\documentclass{jfm}

\usepackage{graphicx}
\usepackage{caption}
\usepackage{amssymb}
\usepackage{amsmath}
\usepackage{natbib}
\usepackage[draft]{hyperref}
\usepackage{wasysym}
\usepackage{dashrule}
\hypersetup{
	colorlinks=true,
	linkcolor=blue,
	citecolor=blue
}
\usepackage[all]{hypcap}

\graphicspath{{./}}

\ifCUPmtlplainloaded \else
\checkfont{eurm10}
\iffontfound
\IfFileExists{upmath.sty}
{\typeout{^^JFound AMS Euler Roman fonts on the system,
		using the 'upmath' package.^^J}%
	\usepackage{upmath}}
{\typeout{^^JFound AMS Euler Roman fonts on the system, but you
		dont seem to have the}%
	\typeout{'upmath' package installed. JFM.cls can take advantage
		of these fonts,^^Jif you use 'upmath' package.^^J}%
}
\else
\fi
\fi

\ifCUPmtlplainloaded \else
\checkfont{msam10}
\iffontfound
\IfFileExists{amssymb.sty}
{\typeout{^^JFound AMS Symbol fonts on the system, using the
		'amssymb' package.^^J}%
	\usepackage{amssymb}%

}{}
\fi
\fi

\ifCUPmtlplainloaded \else
\IfFileExists{amsbsy.sty}
{\typeout{^^JFound the 'amsbsy' package on the system, using it.^^J}%
	\usepackage{amsbsy}}
{}
\fi

\newcommand\Rey{\mbox{\textit{Re}}}  
\newcommand\Pran{\mbox{\textit{Pr}}} 
\newcommand\Ray{\mbox{\textit{Ra}}}  
\newcommand\Nus{\mbox{\textit{Nu}}}  

\newcommand\CFL{\mbox{\textit{CFL}}}

\newsavebox{\astrutbox}
\sbox{\astrutbox}{\rule[-5pt]{0pt}{20pt}}

\newcommand\p		{\ensuremath{\partial}}

\newcommand\eg		{e.g.\,}
\newcommand\cf		{cf. }
\newcommand\ie		{i.e. }

\title[Vertical natural convection: application of the unifying theory of thermal convection]
{Vertical natural convection: application of the unifying theory of thermal convection}

\author[C. S. Ng, A. Ooi, D. Lohse and D. Chung]%
{Chong Shen Ng$^1$%
	\thanks{Email address for correspondence: chongn@unimelb.edu.au},\ns
	Andrew Ooi$^1$, Detlef Lohse$^2$ and Daniel Chung$^1$}

\affiliation{$^1$Department of Mechanical Engineering,
	The University of Melbourne, Victoria 3010, Australia\\[\affilskip]
	$^2$Physics of Fluids Group, Faculty of Science and Technology,
	J.\ M.\ Burgers Center for Fluid Dynamics and MESA+ Institute,
	\\University of Twente, 7500 AE Enschede, The Netherlands}

\date{?; revised ?; accepted ?. - To be entered by editorial office}

\begin{document}

\maketitle

\begin{abstract}
Results from direct numerical simulations of vertical natural convection at
Rayleigh numbers $1.0\times 10^5$--$1.0\times 10^9$ and Prandtl number $0.709$
support a generalised applicability of the Grossmann--Lohse (GL) theory,
which was originally developed for horizontal natural (Rayleigh--B{\'e}nard) convection.
In accordance with the GL theory, it is shown that
the boundary-layer thicknesses of the velocity and
temperature fields in vertical natural convection
obey laminar-like Prandtl--Blasius--Pohlhausen scaling.
Specifically, the normalised mean boundary-layer thicknesses scale with
the $-1/2$-power of a wind-based Reynolds number, where the ``wind''
of the GL theory is interpreted as the maximum mean velocity.
Away from the walls, the dissipation of the turbulent fluctuations,
which can be interpreted as the ``bulk'' or ``background'' dissipation
of the GL theory, is found to obey the Kolmogorov--Obukhov--Corrsin
scaling for fully developed turbulence. In contrast to
Rayleigh--B{\'e}nard convection, the direction
of gravity in vertical natural convection is parallel to the mean flow.
The orientation of this flow presents an added challenge because
there no longer exists an exact relation that links the normalised
global dissipations to the Nusselt, Rayleigh and Prandtl numbers.
Nevertheless, we show that the unclosed term,
namely the global-averaged buoyancy flux that produces the kinetic energy,
also exhibits both laminar  and turbulent scaling behaviours,
consistent with the GL theory.
The present results suggest that, similar
to Rayleigh--B{\'e}nard convection,
a pure power-law relationship between the Nusselt, Rayleigh and
Prandtl numbers is not the best description
for vertical natural convection and existing empirical relationships should
be recalibrated to better reflect the underlying physics.

\end{abstract}

\section{Introduction}

\begin{figure}
\centerline{\includegraphics{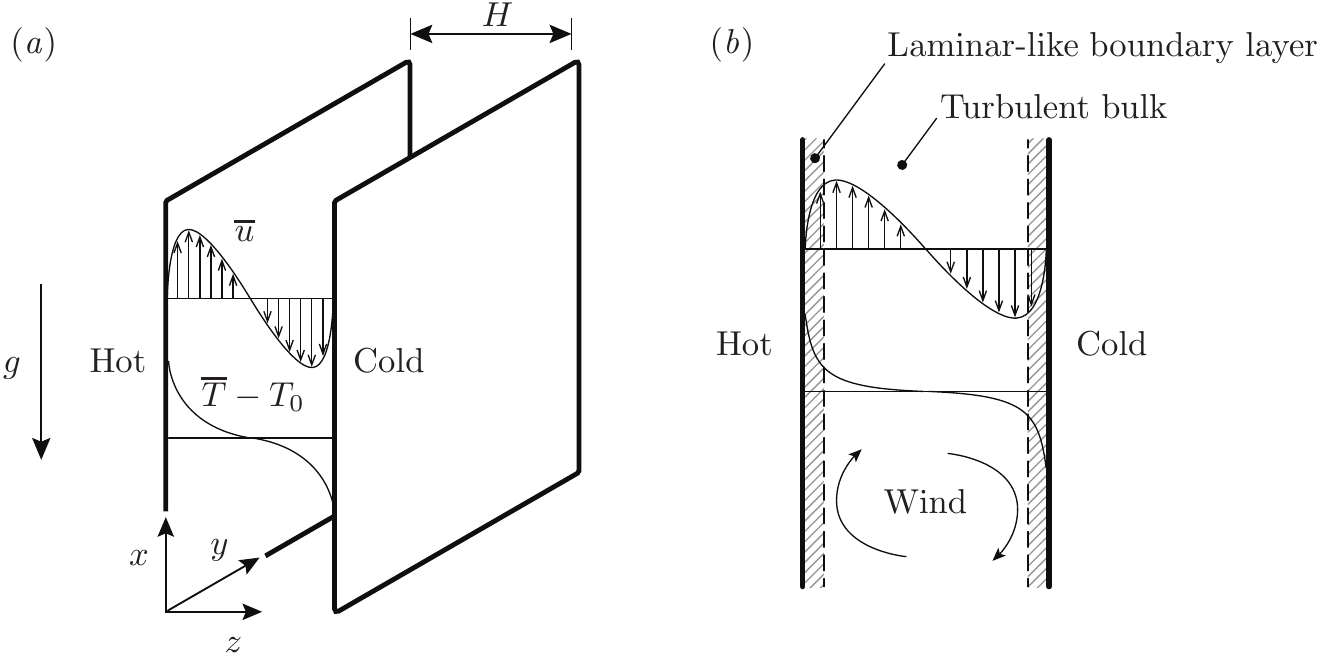}}
\caption{(\textit{a}) Setup of vertical natural convection and
  (\textit{b}) illustration of the laminar-like (boundary-layer) and
  turbulent (bulk) regions of the Grossmann--Lohse theory.}
\label{fig:VerticalChannel}
\end{figure}

\begin{figure}
\centerline{\includegraphics{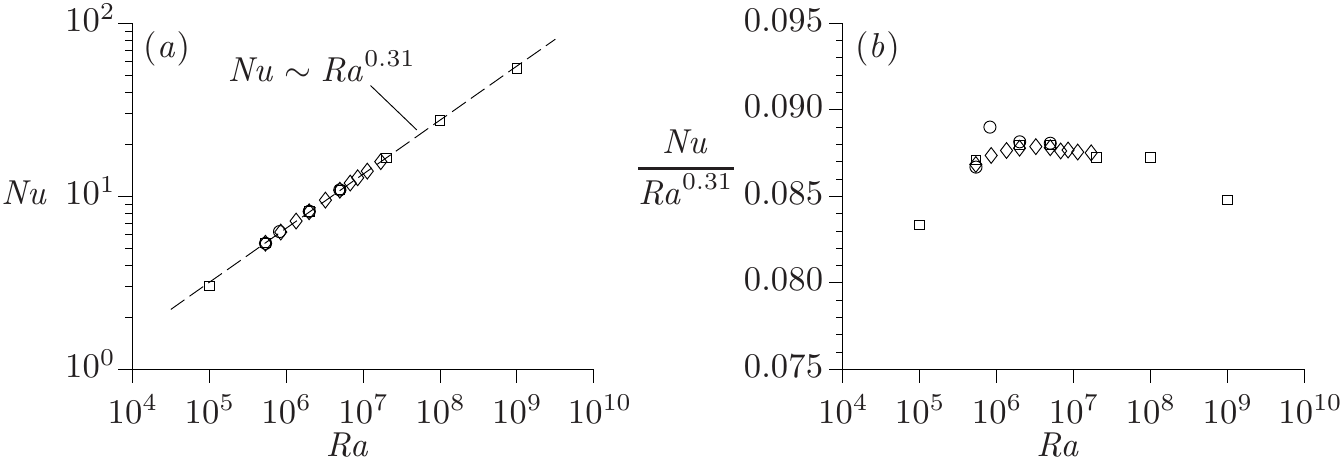}}
\caption{
Trend of $\Nus$ versus $\Ray$ from 
recent DNS data for air ($\Pran = 0.709$):
{\tiny$\square$}, present simulations; 
$\circ$, \citet{Versteegh1999};
{\scriptsize$\lozenge$}, \citet{KisHerwig2012}.
(\textit{a}) $\Nus \sim \Ray^p$, where $p \approx 0.31$
from a least-squares fit to a power law; (\textit{b}) compensated form,
$\Nus/\Ray^p$ versus $\Ray$.
The trend exhibits neither a $1/4$- nor a $1/3$-power scaling.}
\label{fig:NuVsRa}
\end{figure}

In the study of pure buoyancy-driven flow (natural convection)
between two differentially heated vertical surfaces
(figure \ref{fig:VerticalChannel}\,\textit{a}),
there has been
an ongoing interest for establishing a general relationship between the heat
transfer and the temperature difference for an arbitrary fluid.
The heating and cooling that occurs in this vertical setup
is a fundamental problem that is often found in applications
such as building ventilation, computer systems and power plants.
The relevant parameters are the Nusselt number $\Nus$, that is,
the dimensionless heat transfer rate; the Rayleigh number $\Ray$, that is, the
dimensionless temperature difference; and the Prandtl number $\Pran$, that is,
the ratio of fluid viscosity to the thermal diffusivity.
Past studies have shown a preference for the power-law form,
$\Nus \sim\Ray^{p}$ (at fixed $\Pran$), but the exponent $p$ has been
reported to range anywhere between $1/3$ and $1/4$
\citep{Batchelor1954,Elder+Turb.1965,Churchill1975,GeorgeCapp1979,Tsuji1988,
Versteegh1999,KisHerwig2012,Ng2013TurbNatConvectIJHFF}.
A careful examination of recent direct numerical simulation (DNS) data
(figure \ref{fig:NuVsRa}) demonstrates this point:
there is no range in which $\Nus/\Ray^p$
is constant and the effective power-law exponents depend on $\Ray$ and is
less than $1/3$ but greater than $1/4$.
Thus, a pure power law may not be the best
description of the heat-transfer relationship.
One approach is a power-law fit of arbitrary exponent to the
existing data (\eg\,$p \approx 0.31$ in figure \ref{fig:NuVsRa}\,\textit{a}),
but this ignores the underlying flow physics
and is therefore risky when applied outside
the range of calibration.

A similar scaling behaviour has also been reported
in horizontal, \ie Rayleigh--B{\'e}nard (RB), natural convection
\citep[\eg][]{Stevens2011PrRaDependence}.
In RB convection, the unifying theory of \cite{Grossmann+Lohse.2000,Grossmann+Lohse.2001,Grossmann+Lohse.2002,Grossmann+Lohse.2004}
(hereafter GL theory) offered a resolution to the previously
experimentally found \citep{Castaing+etal.1989,Chavanne+etal.1997,Chavanne+authors.2001}
but unexplained $\Nus \sim \Ray^{0.289}$
behaviour (for unity $\Pran$) by showing that the physics-unaware
$0.289$-power can be understood as a combination of a $1/4$- and
a $1/3$-power-law scaling.
The latter two exponents can be readily linked to distinct flow regimes.
The GL theory works because it accounts for the possibility
that, at moderate Rayleigh numbers and away from the walls,
the buoyancy-driven turbulent ``wind'' is not sufficiently strong
to drive a turbulent boundary layer
in the classical sense of Prandtl and von K\'arm\'an.
The theory has since been further articulated and vetted by both experiments
and simulations across a large range of $\Ray$ and $\Pran$
\citep[\eg][]{Ahlers2009heat,Stevens+coauthors.2013}.
The theory has also been extended to other related flows, including
rotating RB convection \citep{Stevens+Clercx+Lohse.2010}
and Taylor--Couette flow \citep{Eckhardt+Grossmann+Lohse.2007}.
The success of the GL theory and the similarities between RB and
vertical natural convection motivates the present study.

In the following, we investigate a generalised application of
the ideas of the GL theory to vertical natural convection
through a close examination of the present DNS data (described in \S\,\ref{sec:dns})
for $\Ray = 1.0\times 10^5$--$1.0\times 10^9$
and $\Pran = 0.709$. Many elements of the GL theory apply to
vertical natural convection. Since the velocity is non-zero in the mean,
the wind of the GL theory is readily identified
and Prandtl--Blasius--Pohlhausen scaling of the boundary layers
is easily verified (\S\,\ref{sec:blscaling}).
The ``bulk'' or ``background'' flow regime (refer to figure \ref{fig:VerticalChannel}\,\textit{b})
described by Kolmogorov--Obukhov--Corrsin scaling is also exhibited by
the dissipation of turbulent fluctuations (\S\,\ref{subsec:BLBulkContributions}).
Apart from the obvious similarities, vertical natural convection
is different to RB convection in one important respect:
the horizontal direction of heat transfer in vertical
natural convection is orthogonal to the vertical direction of
the buoyancy flux, which is the source of turbulent kinetic energy.
The heat flux and the buoyancy flux coincide in RB convection.
Consequently, an exact relationship linking
the global dissipation rate with $\Nus$, $\Ray$ and
$\Pran$ no longer exists (\S\,\ref{subsec:GlobalAveDissip}).
However, it can be shown that the unclosed global-averaged buoyancy flux
also exhibits both laminar and turbulent scaling behaviours, consistent
with the GL theory.
We conclude in \S\,\ref{sec:Conclusions} by summarising current
progress and speculate on future directions towards establishing closure for
a generalised heat-transfer law for vertical natural convection.

\section{Flow setup and direct numerical simulations}
\label{sec:dns}

\subsection{Flow setup}

We adopt the Boussinesq approximation in which density fluctuations
are small relative to the mean.
In this incompressible-flow approximation, the density fluctuation,
which is linearly related to the temperature fluctuation, is dynamically
significant only through the buoyancy force.
The temperature difference, $\Delta T = T_h - T_c$, between the hot
and cold bounding walls drives the fully developed turbulent natural convection
(figure \ref{fig:VerticalChannel}\,\textit{a}). The walls are
separated by the distance $H$.
The governing continuity, momentum and energy equations are
respectively given by,
\begin{subeqnarray}
\dfrac{\p u_i}{\p x_i}	&=&	0,\\ \label{eqn:NonDimContEqn} 
\dfrac{\p u_i}{\p t} + u_j \dfrac{\p u_i}{\p x_j} &=&
		-\frac{1}{\rho_0}\dfrac{\p p}{\p x_i} + \delta_{i1} g \beta (T-T_0) + \nu\dfrac{\p^2 u_i}{\p x_j^2},\\	\label{eqn:NonDimMomEqn} 
\dfrac{\p T}{\p t} + u_j\dfrac{\p T}{\p x_j} &=& \kappa\dfrac{\p^2 T}{\p x_j^2}, \label{eqn:NonDimTempEqn}
\end{subeqnarray}
where $g$ is the gravitational acceleration, $\beta$ is the coefficient
of thermal expansion, $\nu$ is the kinematic viscosity and $\kappa$ is
the thermal diffusivity, all assumed to be independent of temperature.
The coordinate system $x$, $y$ and $z$ (or $x_1$, $x_2$ and $x_3$) refers
to the streamwise (opposing gravity), spanwise and wall-normal directions.
The no-slip and no-penetration boundary conditions are imposed on
the velocity at the walls.
Periodic boundary conditions are imposed on $u_i$, $p$ and $T$ in the
$x$- and $y$-directions.
The Rayleigh, Nusselt and Prandtl numbers are respectively defined by,
\begin{subeqnarray}
\gdef\thesubequation{\theequation \mbox{\textit{a}},\textit{b},\textit{c}}
\Ray \equiv \frac{g\beta\Delta T H^3}{\nu \kappa}, \label{eqn:defineRayleigh}
\quad
\Nus \equiv \frac{f_wH}{\Delta T \kappa}, \label{eqn:defineNusselt}
\quad
\Pran \equiv \frac{\nu}{\kappa}, \label{eqn:definePrandtl}
\end{subeqnarray}
\returnthesubequation
where $f_w \equiv \kappa |\mathrm{d}\overline{T}/\mathrm{d}z|_w$, 
the wall heat flux and $(\cdot)|_w$ denotes the wall value.
Presently, $\overline{(\cdot)}$ denotes the spatial $xy$-plane and
$(\cdot)^\prime$ denotes the corresponding fluctuations.

\subsection{Direct numerical simulations} \label{subsec:DNS}

\begin{table}
	\begin{center}
	\def~{\hphantom{0}}
	\begin{tabular}{ccccccccccc}
	&&&&&& \multicolumn{2}{c}{Wall} & \multicolumn{2}{c}{Centre} & \\ [3pt]
	$\Ray$ 	& $L_x/H$	& $L_y/H$	& $n_x$ & $n_y$ & $n_z$ & $\Delta_{x,y}/\eta$ & $\Delta_z/\eta$  & $\Delta_{x,y}/\eta$  & $\Delta_z/\eta$ & $T_\textit{samp} U_{\Delta T}/H$	\\ [3pt]
	$1.0\times 10^5$	& 8	& 4		& ~384 	& ~192 & ~~96 	& 1.1 & 0.1 & 0.8 &	0.7	& 	1106	\\
	$5.4\times 10^5$	& 8	& 4		& ~384 	& ~192 & ~~96 	& 2.0 & 0.1 & 1.5 &	1.2	& 	1010	\\
	$2.0\times 10^6$	& 8	& 4		& ~384 	& ~192 & ~~96 	& 3.2 & 0.2 & 2.3 &	1.8	&	~862	\\
	$5.0\times 10^6$	& 8	& 4		& ~512 	& ~256 & ~~96 	& 3.3 & 0.3 & 2.4 &	2.5	&	~788	\\
	$2.0\times 10^7$	& 8	& 4		& ~832 	& ~416 & ~192 	& 3.6 & 0.1 & 2.4 &	2.0	&	~802	\\
	$1.0\times 10^8$	& 8	& 4		& 1536 	& ~768 & ~384 	& 3.7 & 0.1 & 2.3 &	1.8	&	~403	\\
	$1.0\times 10^9$	& 8	& 4		& 3200 	& 1600 & ~768 	& 4.5 & 0.1 & 2.6 &	2.1	&	~~~7	\\
	\end{tabular}
	\caption{Simulation parameters of the present DNS cases.}
	\label{tab:SimParam}
	\end{center}
\end{table}

In our simulations, the streamwise, spanwise and wall-normal
domain sizes, $L_x\times L_y\times L_z$, are $8H\times 4H\times H$
and $\Ray=1.0\times 10^5$--$1.0\times 10^9$
(table \ref{tab:SimParam}). The fluid is air with $\Pran =0.709$.
The present grid spacing is uniform in the $x$- and $y$-directions and is stretched
by a cosine map in $z$-direction in order to resolve the steep, near-wall gradients.
The resolutions are chosen so that the simulations resolve the
Kolmogorov scale, $\eta \equiv [\nu^3/\varepsilon_{u^\prime}]^{1/4}$, where
$\varepsilon_{u^\prime}(z) \equiv \nu \overline{(\partial u_i^\prime/\partial x_j)^2}$
is the turbulent dissipation.
In the centre of the channel, $\Delta_{x,y,z} < 2.6\eta$,
while near the wall, $\Delta_{x,y} < 4.5\eta$ and $\Delta_z < 0.3\eta$.
With exception of the highest-$\Ray$ case for which computational
resources are limited, we report statistics averaged over at least
$400$ dimensionless turnover times, where a turnover time is defined
by the free-fall period, $H/U_{\Delta T}$,
where $U_{\Delta T} \equiv (g\beta \Delta T H)^{1/2}$
\citep[\cf][]{Stevens+Verzicco+Lohse.2010}. Higher-$\Ray$ cases are initialised
using interpolated velocity and temperature fields from lower-$\Ray$ cases.
Except for the highest-$\Ray$ case,
the flow is first simulated for more than $70$ dimensionless
turnover times in order to flush out transients
before statistics are sampled. Throughout the sampling duration,
$\Nus$ remains within $5\%$ of its mean,
which is sufficient to ensure a statistically stationary flow
\citep{Stevens+Verzicco+Lohse.2010}.
The switching between exponential growth in $\Nus$ due to
the so-called elevator modes, followed by sudden break-down,
as observed in so-called homogeneous RB
\citep{Calzavarini+Lohse+Toschi+Tripiccione.2005,Calzavarini+authors.2006,Schmidt+authors.2012}
is not observed in the present flow, as there is no
destabilising mean vertical temperature gradient and the flow
is bounded by plates.
The DNS employs a fully conservative fourth-order staggered
finite-difference scheme for the velocity field and the QUICK scheme
to advect the temperature field.
The equations are marched using a low-storage third-order
Runge--Kutta scheme and fractional-step method for enforcing
continuity at $\Delta_t = \CFL\,\max_i (\Delta_i/u_i)$,
where we set $\CFL = \text{1}$
\citep[for details, see][]{Ng2013TurbNatConvectIJHFF,Ng2013MastersThesis}.
A zero-mass-flux constraint is enforced at every time step to
improve convergence, which is similar to using top and bottom
end walls (located far away) in an experiment \citep[\eg][]{Elder+Turb.1965}.

\begin{figure}
\centerline{\includegraphics{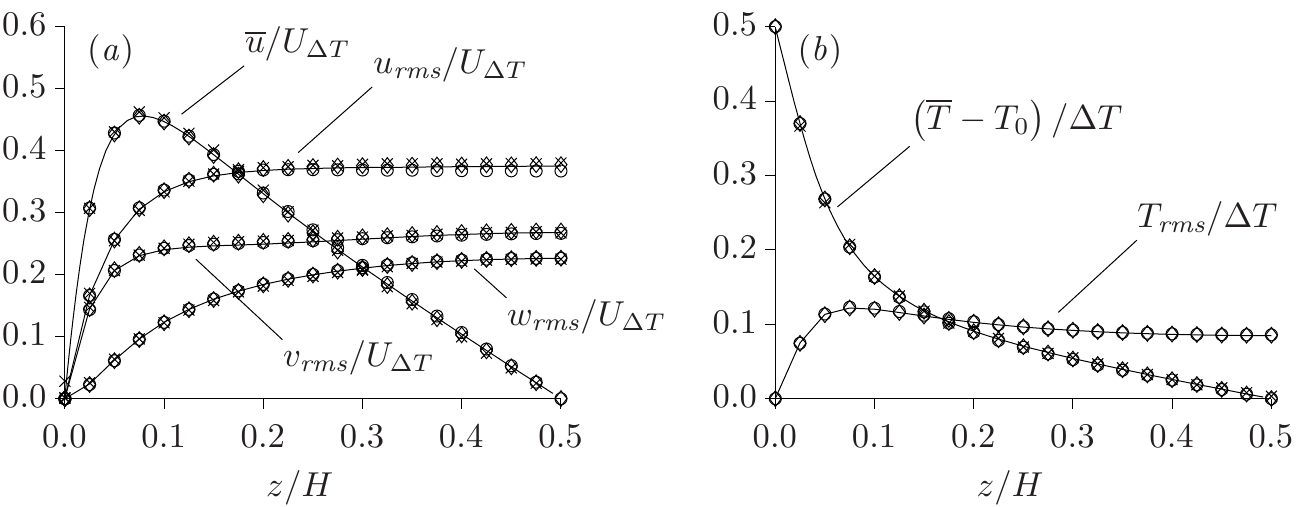}}
\caption{Comparison of mean and second-order
turbulent statistics from DNS for (\textit{a}) velocity and (\textit{b})
temperature, for $\Ray = 5.4\times10^5$:
---, present simulations;
$\circ$, \citet{Versteegh1999};
{\small$\times$}, \citet{Pallares2010};
{\scriptsize$\lozenge$}, \citet{KisHerwig2012}.}
\label{fig:DNSValidation}
\end{figure}

Comparisons of the present simulations with other DNS datasets
\citep{Versteegh1999,Pallares2010,KisHerwig2012} show good
agreement for both mean and second-order statistics
(figure \ref{fig:DNSValidation}).
Throughout this study, statistics are averaged from
both halves of the channel, taking
the antisymmetry (about the centreline) of the mean profiles into account.
The present simulations employ smaller periodic-domain sizes
(two-thirds in each periodic direction) than the
other DNS studies but are chosen in order to resolve
the near-wall region at high $\Ray$.
Simulations conducted with the larger periodic-domain sizes
showed little difference in the mean and second-order statistics,
which are the focus of the present study.

\section{Results and discussion}

The central idea in the GL theory is to conceptually split the flow into
two regions: namely the boundary layer (or plume) and the bulk
(or background) regions
\citep{Grossmann+Lohse.2000,Grossmann+Lohse.2001,Grossmann+Lohse.2004}.
Each of these regions contributes a distinct scaling behaviour to the
total kinetic and thermal dissipations, as discussed in the following.

\subsection{Scaling of boundary-layer thicknesses}
\label{sec:blscaling}

\begin{figure}
\centerline{\includegraphics{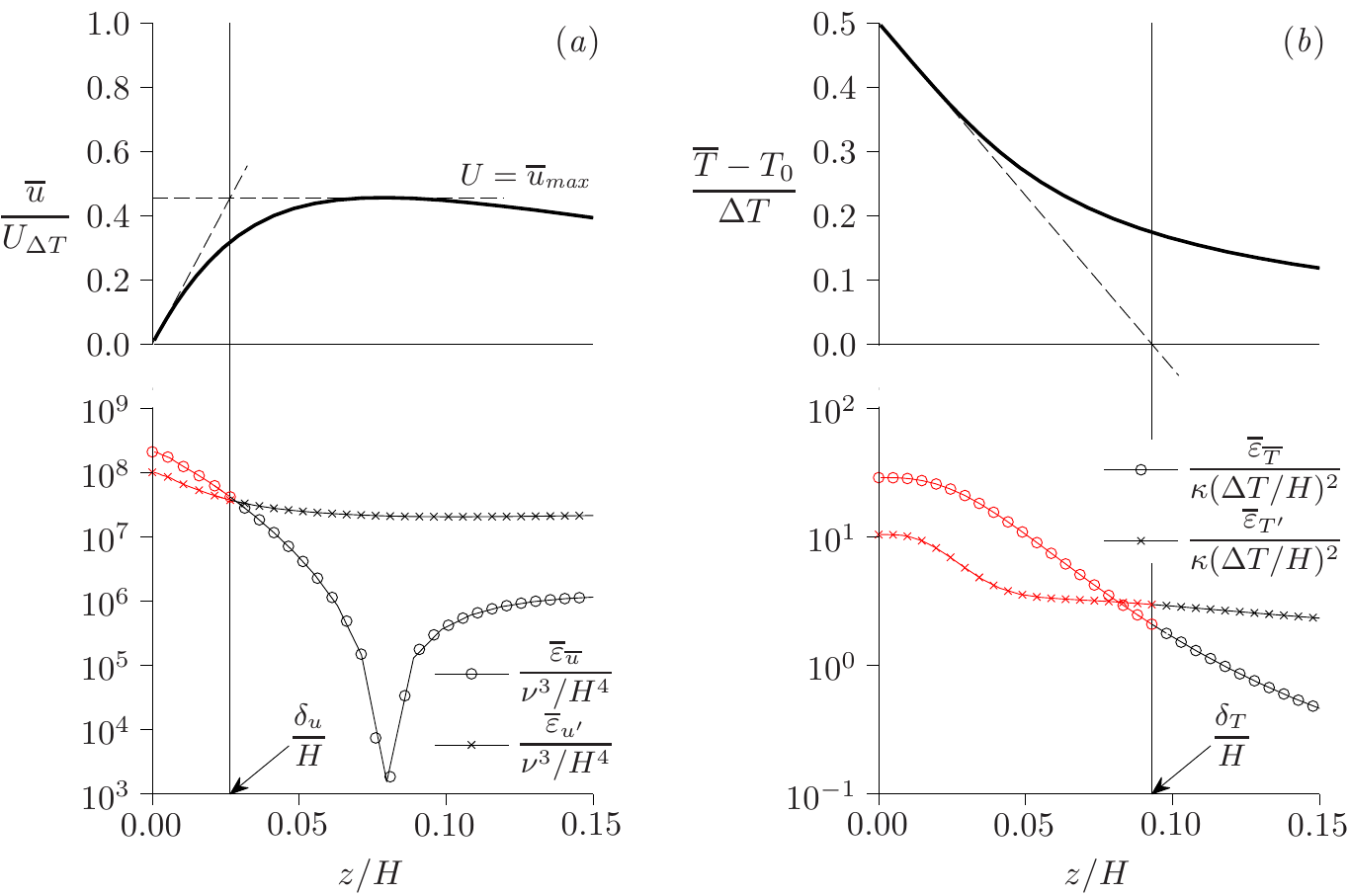}}
\centerline{\includegraphics{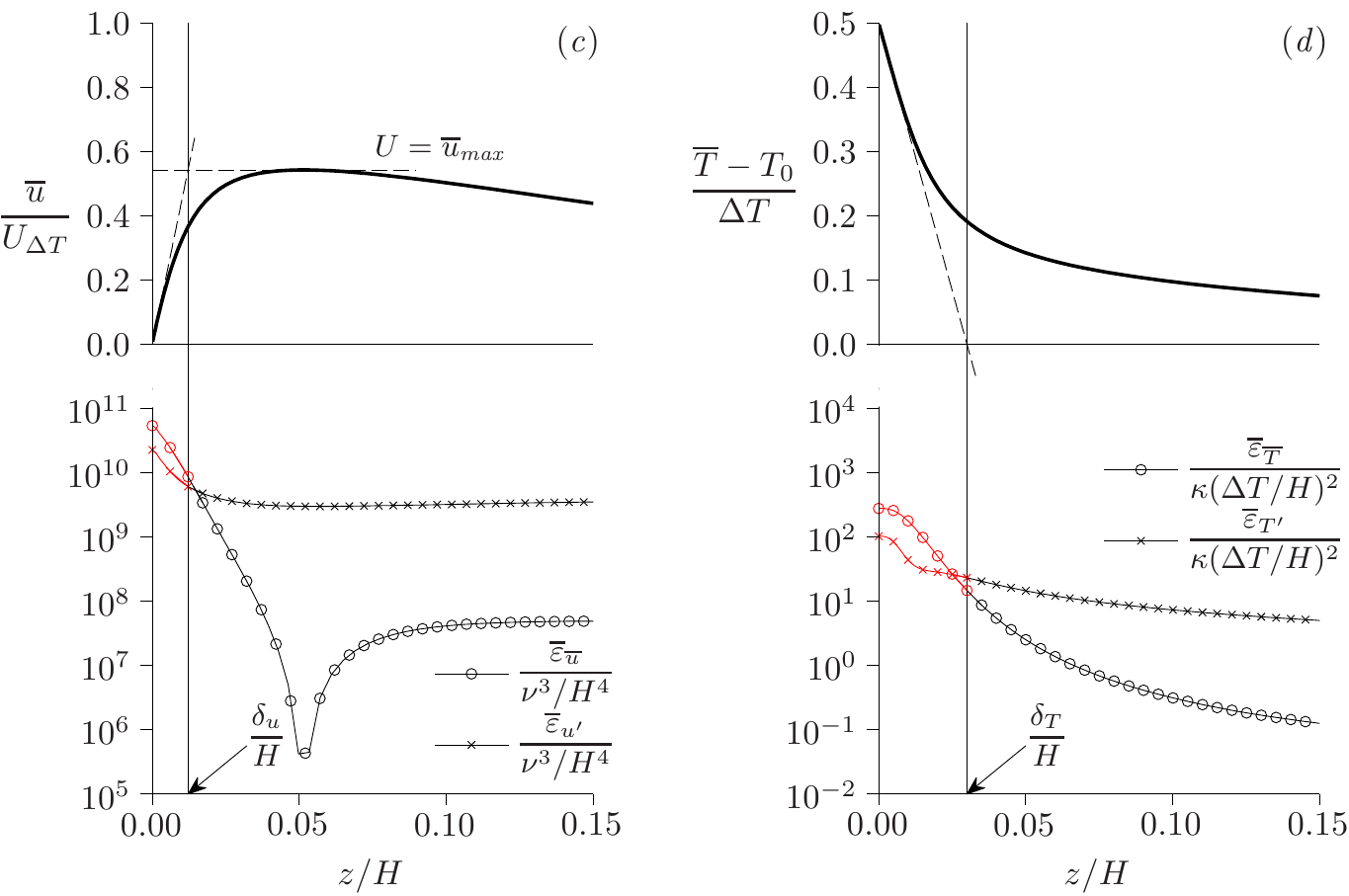}}
\caption{Definitions of the kinetic ($\delta_u$) and thermal ($\delta_T$)
boundary-layer thicknesses shown for DNS data
at $\Ray=5.4\times10^5$ (\textit{a}, \textit{b})
and $\Ray = 2.0\times10^7$ (\textit{c}, \textit{d}).
The kinetic boundary layer is defined as the wall-distance
to the intercept of $\overline{u} = \mathrm{d}\overline{u}/\mathrm{d}z|_w\,z$
and $\overline{u} = \overline{u}_\textit{max}$, and the thermal boundary layer
is defined as the wall-distance to the intercept of
$\overline{T} = T_h + \mathrm{d}\overline{T}/\mathrm{d}z|_w\,z$
and $\overline{T} = T_\textit{h}-\Delta T/2$.
These definitions roughly correspond to the crossover points
between the mean dissipations and turbulent dissipations, \ie
$\overline{\varepsilon}_{\overline{u}}(\delta_u^d)=\overline{\varepsilon}_{u^\prime}(\delta_u^d)$
and $\overline{\varepsilon}_{\overline{T}}(\delta_T^d)=\overline{\varepsilon}_{T^\prime}(\delta_T^d)$.}
\label{fig:definition_deltau_deltaT}
\end{figure}
\begin{figure}
\def\drawline#1#2{\raise 2.5pt\vbox{\hrule width #1pt height #2pt}}
\def\spacce#1{\hskip #1pt}
\centerline{\includegraphics{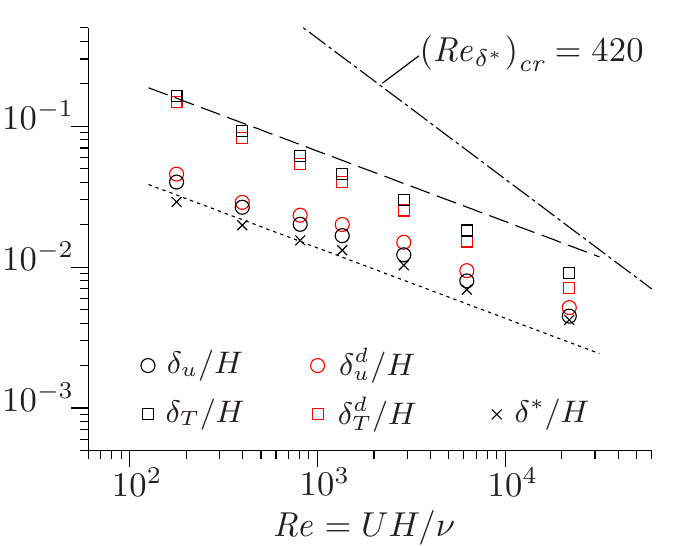}}
\caption{
Trends of normalised boundary-layer thicknesses
appear to scale with the $-1/2$-power law of a
wind-based Reynolds number.
The boundary layer thicknesses are defined as:
the distances to the intercepts (figure \ref{fig:definition_deltau_deltaT}),
$\delta_u/H$ ($\circ$) and
$\delta_T/H$ ({\tiny$\square$}); the crossovers of dissipation
profiles, $\delta^d_u/H$ ({\color{red}$\circ$})
and $\delta^d_T/H$ ({\color{red}\tiny$\square$}); and the displacement
thickness, $\delta^*/H$ ({\scriptsize$\times$}). Shown are the
Prandtl--Blasius--Pohlhausen $-1/2$-power scaling predictions for vertical natural convection (\ref{eqn:AnalyticalPBVNC})
for $\delta_T$ (\hbox{\drawline{4}{0.5}\spacce{2}\drawline{4}{0.5}\spacce{2}\drawline{4}{0.5}\spacce{2}\drawline{2}{0.5}})
and $\delta_u$ (\hbox{\drawline{1.5}{0.5}\spacce{1.5}\drawline{1.5}{0.5}\spacce{1.5}\drawline{1.5}{0.5}\spacce{1.5}
                      \drawline{1.5}{0.5}\spacce{1.5}\drawline{1.5}{0.5}\spacce{1.5}\drawline{1.5}{0.5}\spacce{1.5}\drawline{1.5}{0.5}\spacce{0.5}}).
As reference, the laminar-to-turbulent transition of the shear boundary layer is
expected to occur at $(\Rey_{\delta^*})_\textit{cr} \approx 420$
(\hbox{\drawline{4}{0.5}\spacce{1.5}\drawline{1}{0.5}\spacce{1.5}
       \drawline{4}{0.5}\spacce{1.5}\drawline{1}{0.5}\spacce{1.5}\drawline{4}{0.5}})
\citep{Landau+Lifshitz.1987}.
}
\label{fig:DeltaTrendWithRe}
\end{figure}

For moderate $\Ray$, the GL theory revealed that
the kinetic and thermal boundary-layer thicknesses,
$\delta_u$ and $\delta_T$,
in fact, obey a laminar-like Prandtl--Blasius--Pohlhausen scaling
\citep[\cf][]{Landau+Lifshitz.1987}:
\begin{subeqnarray}\label{eqn:BLdefinitions}
\gdef\thesubequation{\theequation \mbox{\textit{a}},\textit{b},\textit{c}}
\delta_u/H \sim \Rey^{-1/2}, \quad
\delta_T/H \sim \Rey^{-1/2} f(\Pran),\quad
\Rey \equiv U H/\nu,
\end{subeqnarray}
\returnthesubequation
where $U$ refers to the wind.
To test these predictions, we first need to define
$U$, $\delta_u$ and $\delta_T$ for vertical natural convection.
Unlike RB convection where the (mean) streamwise velocity is zero,
the wind is readily identified
for the vertical configuration because of the non-zero persistent
(mean) streamwise velocity (see figure \ref{fig:VerticalChannel}\,\textit{a}).
Presently, it is defined by $U = \overline{u}_\textit{max}$
(figure \ref{fig:definition_deltau_deltaT}\textit{a},\,\textit{c}).
To define $\delta_u$ and $\delta_T$, we
adopt definitions based on the gradient of the time- and plane-averaged velocity
and temperature profiles at the wall \citep[\eg][]{Zhou+Xia.2010,Zhou+etal.2010,Scheel+Schumacher.2014}.
The statistical properties of these definitions were also
first systematically studied by \citet{Sun+Cheung+Xia.2008}.
For the hot wall, the kinetic boundary-layer thickness, $\delta_u$, is defined as the
wall-normal distance to the intercept of
$\overline{u} = \mathrm{d}\overline{u}/\mathrm{d}z|_w\,z$ and $\overline{u} = U$
(figure \ref{fig:definition_deltau_deltaT}\textit{a},\,\textit{c}),
\ie $\delta_u = U/(\mathrm{d} \overline{u}/\mathrm{d}z|_w)$,
while the thermal boundary-layer thickness,
$\delta_{T}$ is defined as the wall-normal distance to the intercept of
$\overline{T} = T_h + \mathrm{d}\overline{T}/\mathrm{d}z|_w\,z$ and
$\overline{T} = T_h - \Delta T/2$
(figure \ref{fig:definition_deltau_deltaT}\textit{b},\,\textit{d}),
\ie $\delta_T = -(\Delta T/2)/(\mathrm{d} \overline{T}/\mathrm{d}z|_w)$.
These boundary-layer definitions conveniently
distinguish the boundary-layer behaviour of the flow from the bulk
behaviour and this is demonstrated for two representative $\Ray$
in figure \ref{fig:definition_deltau_deltaT}.
In figure \ref{fig:definition_deltau_deltaT}(\textit{a},\,\textit{c}),
the kinetic dissipation due to the mean, $\overline{\varepsilon}_{\overline{u}}
 \equiv \nu \overline{(\partial \overline{u}_i/\partial x_j)^2}
= \nu (\mathrm{d}\overline{u}/\mathrm{d} z)^2$, overwhelms
the kinetic dissipation due to the turbulent fluctuations,
$\overline{\varepsilon}_{u^\prime} \equiv \nu \overline{(\p u_i^\prime/\p x_j)^2}$
in the kinetic boundary layer. Similarly, in figure
\ref{fig:definition_deltau_deltaT}(\textit{b},\,\textit{d}),
the thermal dissipation due to the mean,
$\overline{\varepsilon}_{\overline{T}}  \equiv \kappa \overline{(\partial \overline{T}/\partial x_j)^2}
= \kappa (\mathrm{d}\overline{T}/\mathrm{d} z)^2$,
overwhelms the thermal dissipation due to the turbulent fluctuations,
$\overline{\varepsilon}_{T^\prime} \equiv  \kappa \overline{(\p T^\prime/\p x_j)^2}$,
in the thermal boundary layer.
Both profiles of $\overline{\varepsilon}_{u^\prime}$ and
$\overline{\varepsilon}_{T^\prime}$ exhibit
characteristics similar to that found in RB convection:
the profiles peak at the wall and
are approximately flat in the bulk
\citep[\eg][]{Emran+Schumacher.2008,Kaczorowski+Wagner.2009,Kaczorowski+Xia.2013}.
Alternative boundary-layer definitions such as the crossover locations
between the mean dissipations and fluctuation dissipations,
$\delta_u^d$ and $\delta_T^d$, as well as the displacement thickness,
$\delta^* \equiv \int_{0}^{\delta_\textit{max}} (1 - \overline{u}/\overline{u}_\textit{max})\,\mathrm{d}z$,
where $\overline{u}(\delta_\textit{max}) = \overline{u}_\textit{max}$,
are found to provide similar scaling
characteristics, as verified in figure \ref{fig:DeltaTrendWithRe}.

For comparison, we compute the
Prandtl--Blasius--Pohlhausen boundary-layer thicknesses
for vertical natural convection from the laminar similarity
scaling, which is different to its horizontal counterpart.
Using the definitions for $\delta_u$, $\delta_T$ (figure
\ref{fig:definition_deltau_deltaT}) and wind-based
$\Rey$ from (\ref{eqn:BLdefinitions}\,\textit{c}) and for $\Pran = 0.709$,
we obtain, by setting $x/H = 1$
in the laminar similarity scaling \cite[see][\S\,4-13.3]{White.1991}:
\begin{subeqnarray}\label{eqn:AnalyticalPBVNC}
\gdef\thesubequation{\theequation \mbox{\textit{a}},\textit{b}}
\delta_u/H
 \approx 0.43\,\Rey^{-1/2}, \quad
\delta_T/H
 \approx 2.10\,\Rey^{-1/2}.
\end{subeqnarray}
\returnthesubequation
Varying $x/H$, pertaining to the wall-parallel coherence of the wind,
would merely alter the coefficients in (\ref{eqn:AnalyticalPBVNC}).
In figure \ref{fig:DeltaTrendWithRe}, the boundary-layer thicknesses
using the slope definition from figure \ref{fig:definition_deltau_deltaT},
\ie $\delta_u$ and $\delta_T$, the dissipation crossover definitions,
$\delta^d_u$ and $\delta^d_T$, and displacement thickness $\delta^*$,
are compared with (\ref{eqn:AnalyticalPBVNC}\textit{a},\,\textit{b}).
Using a least-squares fit of the present data to a power law,
we find that $\delta_u/H \sim \Rey^{-0.45}$ and $\delta_T/H \sim \Rey^{-0.60}$
(not shown in figure \ref{fig:DeltaTrendWithRe})
which is in fair agreement to the $\Rey^{-1/2}$ trend,
in accordance to the laminar
predictions from the GL theory.
Hence, for simplicity, we will adopt the boundary-layer
definitions based on $\delta_u$ and $\delta_T$ hereafter.
An upper bound for the boundary layers can be obtained
when both boundary-layer and bulk regions are laminar.
In this case, the velocity profile is a cubic and the temperature profile
is linear, from which
$\delta_u/H \approx 0.096$ and $\delta_T/H = 0.5$.
For reference, the laminar-to-turbulent transition which
occurs at $(\Rey_{\delta^*})_\textit{cr} \equiv (U \delta^*/\nu)_\textit{cr} \approx 420$,
where $\delta^*$ is the displacement thickness
\citep{Landau+Lifshitz.1987}, is also shown in
figure \ref{fig:DeltaTrendWithRe}, to the right of all present data.
Consistent with the insight provided by the GL theory,
the boundary layers in vertical natural convection for
the present $\Ray$ range cannot be considered
as turbulent boundary layers.
Instead, they can be interpreted as laminar boundary
layers animated by the turbulent wind.

Figure \ref{fig:DeltaTrendWithRe} shows that $\delta_T > \delta_u$
in all cases considered here at $\Pran = 0.709$.
This situation is expected to be reversed ($\delta_u > \delta_T$) when $\Pran > 1$
\citep{Grossmann+Lohse.2001}. At transitional $\Ray$
and at high $\Pran$, an oscillatory flow regime is found in
vertical natural convection \citep{Chait+Korpela.1989} and it remains
unknown whether this oscillatory flow persists
at higher $\Ray$ and whether (\ref{eqn:BLdefinitions}\,\textit{b})
accounts for this behaviour.

\subsection{Boundary-layer and bulk contributions to the dissipations} \label{subsec:BLBulkContributions}
\begin{figure}
\centerline{\includegraphics{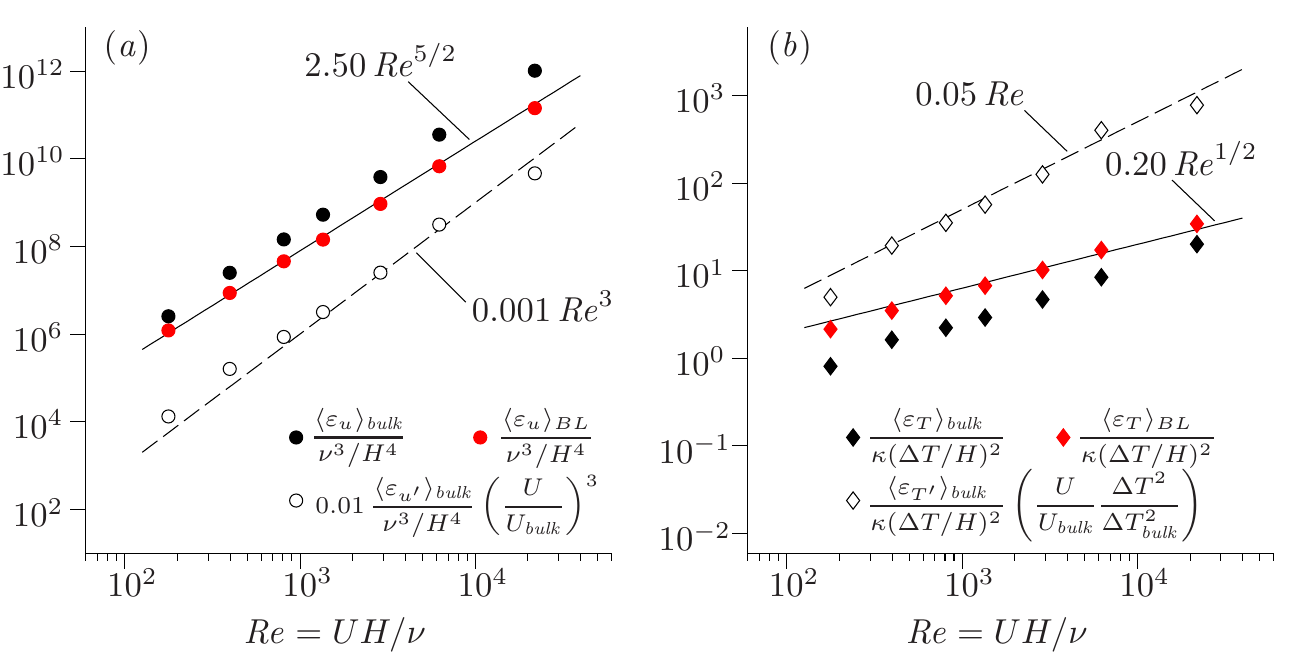}}
\caption{Dissipation trends in the boundary layer and bulk for
(\textit{a}) $\langle \varepsilon_u \rangle$ and (\textit{b}) $\langle \varepsilon_T \rangle$.
The figures show that $\langle \varepsilon_{u} \rangle_{BL}\sim \Rey^{5/2}$,
whilst $\langle \varepsilon_{T} \rangle_\textit{bulk} \sim \langle \varepsilon_{T} \rangle_{BL} \sim \Rey^{1/2}$.
Also shown are bulk dissipations of turbulent fluctuations which vary as
$\langle \varepsilon_{u^\prime} \rangle_\textit{bulk} \sim \Rey^{3}$ ($\circ$)
and $\langle \varepsilon_{T^\prime} \rangle_\textit{bulk} \sim \Rey$ ({\scriptsize$\Diamond$}).}
\label{fig:epsversusRe}
\end{figure}

The GL theory splits the  global-averaged kinetic and thermal dissipation rates
into contributions from the boundary layer and bulk regions (figure \ref{fig:VerticalChannel}) such that
\begin{subeqnarray} \label{DissipationSplit.GLTheory}
\langle \varepsilon_u \rangle
		&=& \langle \varepsilon_{u} \rangle_{BL} + \langle \varepsilon_{u} \rangle_\textit{bulk}
		 = \dfrac{2}{H}\int_{0}^{\delta_u}\nu\left(\dfrac{\p u_i}{\p x_j}\right)^2\mathrm{d}z 
		 	+ \dfrac{2}{H}\int_{\delta_u}^{H/2}\nu\left(\dfrac{\p u_i}{\p x_j}\right)^2\mathrm{d}z, \label{DissipationSplitU.GLTheory} \\
\langle \varepsilon_T \rangle 
		&=& \langle \varepsilon_{T} \rangle_{BL} + \langle \varepsilon_{T} \rangle_\textit{bulk}
		 = \dfrac{2}{H}\int_{0}^{\delta_T}\kappa\left(\dfrac{\p T}{\p x_j}\right)^2\mathrm{d}z 
		 	+ \dfrac{2}{H}\int_{\delta_T}^{H/2}\kappa\left(\dfrac{\p T}{\p x_j}\right)^2\mathrm{d}z, \label{DissipationSplitT.GLTheory}
\end{subeqnarray}
\cf (2.9) and (2.10) in \cite{Grossmann+Lohse.2000}, where the time- and
volume-average is denoted by $\langle \cdot \rangle$.
Following the GL theory, once the wind that acts on the boundary layer is
identified as $U$, the kinetic boundary-layer dissipation, $\langle \varepsilon_{u} \rangle_{BL}$,
is approximated using the wall-normal gradient of the streamwise velocity,
\ie $\nu(\p u_i/\p x_j)^2 \approx \nu(U/\delta_u)^2$, over the volume fraction, $\delta_u/H$.
Similarly, the thermal boundary-layer dissipation, $\langle \varepsilon_{T} \rangle_{BL}$,
is approximated using the wall-normal gradient of the temperature,
\ie $\kappa(\p T/\p x_j)^2 \approx \kappa(\Delta T/\delta_T)^2$, over the volume-fraction $\delta_T/H$.
The boundary-layer terms on the right-hand side of
(\ref{DissipationSplit.GLTheory}\textit{a},\,\textit{b}) can thus be written as,
\begin{subeqnarray} \label{eqn:eps_BL}
\langle \varepsilon_{u} \rangle_{BL} 
	&\sim& \nu \frac{U^2}{\delta_u^2}\left(\frac{\delta_u}{H}\right)
	 \sim \nu \frac{U^2}{\delta_u^2}\left(\Rey^{-1/2}\right)
	 = \frac{\nu^3}{H^4}\Rey^{5/2}, \\
\langle \varepsilon_{T} \rangle_{BL} 	
	&\sim& \kappa \frac{\Delta T^2}{\delta_T^2}\left(\frac{\delta_T}{H}\right)
	 \sim \kappa \frac{\Delta T^2}{\delta_T^2}\left(\Rey^{-1/2} f(\Pran)\right)
	 = \kappa\dfrac{\Delta T^2}{H^2}\Rey^{1/2} f(\Pran),
\end{subeqnarray}
where the expressions for $\delta_u/H$ and $\delta_T/H$ in
(\ref{eqn:BLdefinitions}\textit{a},\,\textit{b}) are used.
On the other hand, the bulk dissipation terms are modelled as
\begin{subeqnarray}\label{eqn:eps_bulk}
\gdef\thesubequation{\theequation \mbox{\textit{a}},\textit{b}}
\langle \varepsilon_{u} \rangle_\textit{bulk} 
	\sim \dfrac{U^3}{H} 
	= \dfrac{\nu^3}{H^4} \Rey^3, \qquad
\langle \varepsilon_{T} \rangle_\textit{bulk} 
	\sim \dfrac{U\Delta T^2}{H} 
	= \kappa \dfrac{\Delta T^2}{H^2}\Pran \Rey,
\end{subeqnarray}
\returnthesubequation
which follow from dimensional arguments of the turbulence cascade
in the bulk region. In this region,
larger eddies transfer energy to smaller eddies. Thus,
the dissipation rate can be thought to scale with
the largest eddies with energy of order $U^2$
and timescale $H/U$, independent of $\nu$. Similarly,
the thermal dissipation rate can be thought to scale with the
largest eddies with variance of order $\Delta T^2$ and
timescale $H/U$, independent of $\kappa$ \citep[see][]{Pope2000turbulent}.
Figure \ref{fig:epsversusRe}(\textit{a},\,\textit{b}) show the trends of the
boundary-layer and bulk contributions. In figure \ref{fig:epsversusRe}\,(\textit{a}),
although $\langle \varepsilon_{u} \rangle_{BL} \sim \Rey^{5/2}$ and
$\langle \varepsilon_{u} \rangle_\textit{bulk} \sim \Rey^3$
as predicted in (\ref{eqn:eps_BL}\,\textit{a}) and (\ref{eqn:eps_bulk}\,\textit{a}),
the ratio of boundary-layer-to-bulk contributions for thermal dissipation appears
constant as shown by the parallel trends of $\langle \varepsilon_{T} \rangle_{BL}$
and $\langle \varepsilon_{T} \rangle_\textit{bulk}$ in figure \ref{fig:epsversusRe}\,(\textit{b}).
This seemingly contradicts the $\langle \varepsilon_{T} \rangle_{BL} \sim \Rey^{1/2}$ and
$\langle \varepsilon_{T} \rangle_\textit{bulk}\sim \Rey$ predictions for the
boundary-layer and bulk thermal dissipations, (\ref{eqn:eps_BL}\,\textit{b})
and (\ref{eqn:eps_bulk}\,\textit{b}). A similar behaviour is reported by
\citet{Grossmann+Lohse.2004} based on a DNS study of RB convection by
\citet{Verzicco+Camussi.2003}. The reason is that plumes, which provide the scaling in
(\ref{eqn:eps_BL}\,\textit{b}), are also present in the bulk, as discussed in
\citet{Grossmann+Lohse.2004}.

\begin{figure}
\centerline{\includegraphics{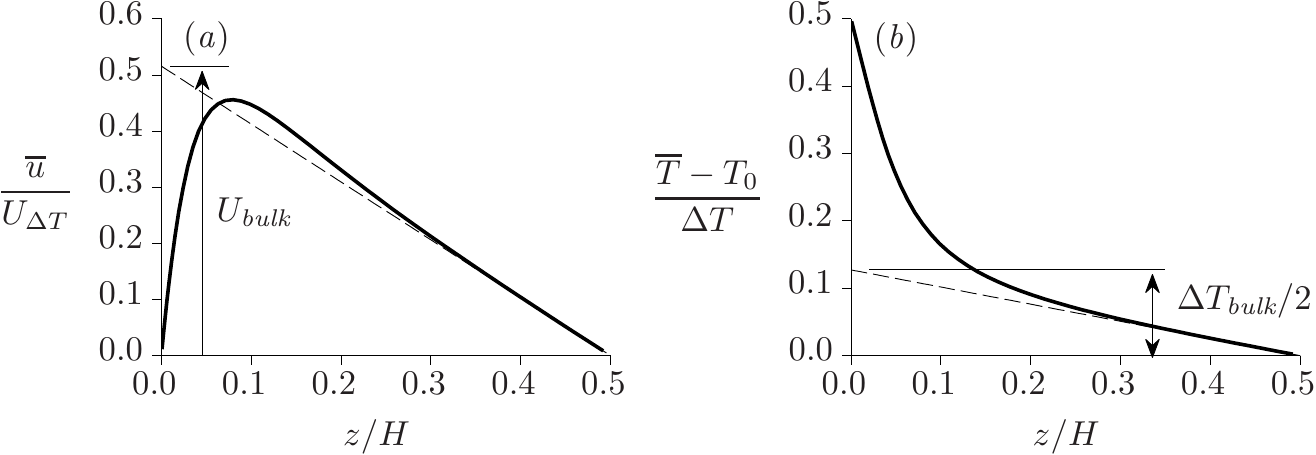}}
\caption{Illustrations of (\textit{a}) $U_\textit{bulk}$ and
(\textit{b}) $\Delta T_\textit{bulk}$ for $\Ray = 5.4\times10^5$.
Specifically, they are defined as
$U_\textit{bulk} = -(H/2)\,\mathrm{d}\overline{u}/\mathrm{d}z|_c$
and $\Delta T_\textit{bulk} = -H\,\mathrm{d}\overline{T}/\mathrm{d} z|_c$,
where $(\cdot)|_c$ denotes the centreline value,
refer to (\ref{eqn:BulkBLDefinitions}).}
\label{fig:BulkBLDefinitions} 
\end{figure}
It seems unexpected that the classical cascade arguments that lead to the $\Rey$ scaling
for $\varepsilon_{T,\textit{bulk}}$ are not observed in the present flow.
Presently, we consider the possibility that the turbulent scalings in the
bulk are obscured by a strong mean component. To observe this behaviour,
we subtract the bulk dissipation of the mean,
\begin{subeqnarray}\label{eqn:IsolateEpsFluct}
\gdef\thesubequation{\theequation \mbox{\textit{a}},\textit{b}}
\langle \varepsilon_{u^\prime} \rangle_\textit{bulk} 
	= \langle \varepsilon_{u} \rangle_\textit{bulk} - \langle \varepsilon_{\overline{u}} \rangle_\textit{bulk}\,, \quad
\langle \varepsilon_{T^\prime} \rangle_\textit{bulk} 
	= \langle \varepsilon_{T} \rangle_\textit{bulk} - \langle \varepsilon_{\overline{T}} \rangle_\textit{bulk}\,,
\end{subeqnarray}
\returnthesubequation
where,
\begin{subeqnarray}\label{eqn:EpsBulkMean.Definition}
\gdef\thesubequation{\theequation \mbox{\textit{a}},\textit{b}}
\langle \varepsilon_{\overline{u}} \rangle_\textit{bulk} =  
	\dfrac{2}{H}\int_{\delta_u}^{H/2}\nu \left(\dfrac{\mathrm{d} \overline{u}}{\mathrm{d} z}\right)^2\mathrm{d}z, \quad
\langle \varepsilon_{\overline{T}} \rangle_\textit{bulk} =  
	\dfrac{2}{H}\int_{\delta_T}^{H/2}\kappa\left(\dfrac{\mathrm{d}\overline{T}}{\mathrm{d}z}\right)^2\mathrm{d}z.
\end{subeqnarray}
\returnthesubequation
For RB convection, the global average of (\ref{eqn:EpsBulkMean.Definition}\,\textit{a})
is zero although (\ref{eqn:EpsBulkMean.Definition}\,\textit{b}) is non-zero.
Indeed, it will be shown that the strong mean components in vertical natural
convection, $\Delta T_\textit{bulk}$ and $U_\textit{bulk}$, drive the turbulent fluctuations,
as discussed in \citet{Grossmann+Lohse.2004} in the context of RB convection.
Here, we define $\Delta T_\textit{bulk}$ and $U_\textit{bulk}$ using their
corresponding centreline mean gradients (figure \ref{fig:BulkBLDefinitions}),
\begin{subeqnarray}\label{eqn:BulkBLDefinitions}
\gdef\thesubequation{\theequation \mbox{\textit{a}},\textit{b}}
\Delta T_\textit{bulk}
	=  -H \dfrac{\mathrm{d}\overline{T}}{\mathrm{d} z}\biggr|_c, \qquad
U_\textit{bulk}
	=  -\dfrac{H}{2} \dfrac{\mathrm{d}\overline{u}}{\mathrm{d}z}\biggr|_c,
\end{subeqnarray}
\returnthesubequation
where $(\cdot)|_c$ denotes the centreline value.
Thus, the bulk dissipations due to fluctuating quantities may now scale as
\begin{subeqnarray} \label{eqn:epsFluctBulk}
\langle \varepsilon_{u^\prime} \rangle_\textit{bulk}  
	&\sim&  \dfrac{U^3_\textit{bulk}}{H} 
	 = \dfrac{\nu^3}{H^4} \Rey^3 \left(\dfrac{U_\textit{bulk}}{U}\right)^3, \\ 
\langle \varepsilon_{T^\prime} \rangle_\textit{bulk} 
	&\sim&  \dfrac{U_\textit{bulk}\,\Delta T^2_\textit{bulk}}{H} 
	 = \kappa \dfrac{\Delta T^2}{H^2}\Pran \Rey \left(\dfrac{U_\textit{bulk}}{U}\dfrac{\Delta T^2_\textit{bulk}}{\Delta T^2}\right),
\end{subeqnarray}
where the wind-based Reynolds number scaling, $\Rey$, is defined as before.
In figure \ref{fig:epsversusRe}(\textit{a},\,\textit{b}), we find that the
trends predicted by (\ref{eqn:epsFluctBulk}) for $\langle \varepsilon_{u^\prime} \rangle_\textit{bulk}$
and $\langle \varepsilon_{T^\prime} \rangle_\textit{bulk}$ agree with the
power-laws of the GL theory for bulk dissipation (\ref{eqn:eps_bulk}),
and are consistent with the Kolmogorov--Obukhov--Corrsin scaling in the bulk region.
Thus, to fully extend the GL theory to the present flow,
(\ref{eqn:EpsBulkMean.Definition}) and (\ref{eqn:epsFluctBulk}) need to be closed
with models for  $U_\textit{bulk}/U$, $\Delta T_\textit{bulk}/\Delta T$
and the bulk dissipation of the mean, \ie $\langle \varepsilon_{\overline{u}} \rangle_\textit{bulk}$
and $\langle \varepsilon_{\overline{T}} \rangle_\textit{bulk}$,
in terms of $\Rey$, $\Ray$, $\Nus$ and $\Pran$.

\subsection{Global averages for kinetic and thermal dissipations} \label{subsec:GlobalAveDissip}
\begin{figure}
\centerline{\includegraphics{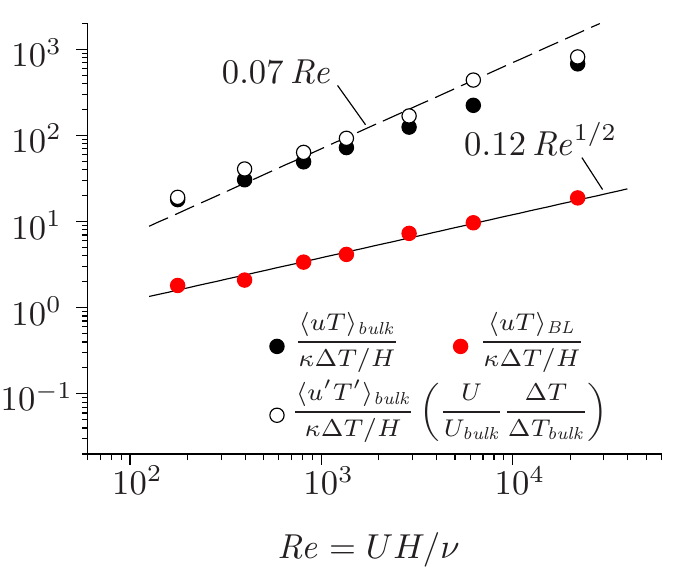}}
\caption{Trends of the buoyancy flux, $\langle uT \rangle$, showing
$\langle uT \rangle_\textit{BL} \sim \Rey^{1/2}$  and
$\langle u^\prime T^\prime \rangle_\textit{bulk} \sim \Rey$.
The boundary-layer and bulk components are decomposed
using $\delta_u = U/(\mathrm{d} \overline{u}/\mathrm{d}z|_w)$, as before.}
\label{fig:BuoyIntegralVsRe}
\end{figure}
For both RB and vertical natural convection, the global-averaged dissipations rates
in (\ref{DissipationSplitU.GLTheory}) take the exact forms,
\begin{subeqnarray}\label{eqn:GlobalDissipation}
\gdef\thesubequation{\theequation \mbox{\textit{a}},\textit{b}}
\langle \varepsilon_u \rangle 
	&=  \dfrac{\nu^3}{H^4} \dfrac{\langle -u_g T \rangle}{\kappa \Delta T/H} \dfrac{\Ray}{\Pran^{2}}, \qquad
\langle \varepsilon_T \rangle 
	&=  \kappa \dfrac{\Delta T^2}{H^2} \Nus,
\end{subeqnarray}
\returnthesubequation
where $u_g$ is velocity component in the direction of gravity.
In RB convection, $\langle -u_{g} T \rangle = \langle w T \rangle = f_w - \langle \kappa\,\mathrm{d}\overline{T}/\mathrm{d}z \rangle$, and it can thus be
shown that
$\langle \epsilon_u \rangle_{RB} = (\nu^3/H^4)(\Nus - 1) (\Ray/\Pran^2)$,
\cf (9) and (10) in \citet{Ahlers2009heat}.
In contrast, $\langle -u_g T \rangle = \langle u T \rangle$ for
vertical natural convection and the relations remain unclosed. However, if we
apply the same GL-theory scaling arguments for boundary-layer contribution
\ie $\langle uT \rangle_\textit{BL}$,
and the same GL-theory scaling arguments for the turbulent bulk contribution,
\ie $ \langle u^\prime T^\prime \rangle_\textit{bulk} = \langle u T \rangle_\textit{bulk} - \langle \overline{u}\overline{T} \rangle_\textit{bulk} $,
as before, we obtain,
\begin{subeqnarray} \label{eqn:VerticalHeatFluxScaling}
\langle u T \rangle_\textit{BL} &\sim&  U\Delta T \dfrac{\delta_u}{H}
	= \kappa \dfrac{\Delta T}{H} \Pran \Rey \left(\dfrac{\delta_u}{H}\right)
	\sim  \kappa \dfrac{\Delta T}{H} \Pran \Rey^{1/2}, \\
\langle u^\prime T^\prime \rangle_\textit{bulk} &\sim& U_\textit{bulk} \Delta T_\textit{bulk}
	= \kappa \dfrac{\Delta T}{H} \Pran \Rey
	\left(\dfrac{U_\textit{bulk}}{U}\dfrac{\Delta T_\textit{bulk}}{\Delta T}\right).
\end{subeqnarray}
In figure \ref{fig:BuoyIntegralVsRe}, we find that
$\langle u T \rangle_\textit{BL} \sim \Rey^{1/2}$ and 
$\langle u^\prime T^\prime \rangle_\textit{bulk} \sim \Rey$,
in agreement with (\ref{eqn:VerticalHeatFluxScaling}) and
corroborating the GL theory of differing physics
in the boundary layer and bulk regions.
Similar to the thermal dissipation discussed in \S\,\ref{subsec:BLBulkContributions}, the contamination of the bulk region by
plumes released from the laminar boundary layer \citep{Grossmann+Lohse.2004}
results in a scaling exponent for the bulk
contribution that is less than $1$ but larger than $1/2$
(compare figures \ref{fig:epsversusRe}\,\textit{b} and \ref{fig:BuoyIntegralVsRe}).
This suggests a possible approach for modelling the unclosed buoyancy
flux (\ref{eqn:GlobalDissipation}\,\textit{a}) once
appropriate models can be found for $U_\textit{bulk}/U$, $\Delta T_\textit{bulk}/\Delta T$
and the mean component of the buoyancy flux,
\ie $\langle \overline{u}\overline{T} \rangle_\textit{bulk}$.

\section{Conclusions} \label{sec:Conclusions}
The present DNS data for vertical natural convection with
$\Ray$ ranging between $1.0\times 10^5$ and $1.0\times 10^9$ and
$\Pran = 0.709$ demonstrate the general applicability of the GL
theory which was originally developed for RB convection. In agreement
with the theory, the $\Nus\sim\Ray^p$ relationship for vertical
natural convection exhibits neither a $1/3$- nor a $1/4$-power scaling
due to the different physics of the boundary layer (or plume) and bulk
(or background). Thus, the dissipation in the boundary layer and bulk,
(\ref{DissipationSplitU.GLTheory}), are expected to scale differently
as proposed by the GL theory. Similar to RB convection, the boundary-layer
thicknesses of velocity and temperature for vertical natural convection
exhibit laminar-like scaling, \ie $\delta_u/H \sim \Rey^{-1/2}$ and
$\delta_T/H \sim \Rey^{-1/2}$ (figure \ref{fig:DeltaTrendWithRe}),
where the wind-based Reynolds number is defined as $\Rey \equiv U H/\nu$.
For the present configuration, the ``wind'' is readily identified from
the non-zero plane-averaged streamwise velocity,
$U = \overline{u}_\textit{max}$.
In the boundary layers, the kinetic and thermal dissipations scale as
predicted by the GL theory, (\ref{eqn:eps_BL}), \ie $\langle \varepsilon_{u} \rangle_{BL} \sim \Rey^{5/2}$
and $\langle \varepsilon_{T} \rangle_{BL} \sim \Rey^{1/2}$ (figure \ref{fig:epsversusRe}).
In the bulk region, the Kolmogorov--Obukhov--Corrsin scaling, 
\ie $\langle \varepsilon_{u^\prime} \rangle_\textit{bulk} \sim \Rey^{3}$
and $\langle \varepsilon_{T^\prime} \rangle_\textit{bulk} \sim \Rey$,
are recovered once the dissipations of the mean are subtracted
from the bulk dissipations (figure \ref{fig:epsversusRe}).
These are consistent with the power laws originally predicted by the GL theory, (\ref{eqn:eps_bulk}).
Unlike RB convection, the global kinetic dissipation (\ref{eqn:GlobalDissipation}\,\textit{a})
cannot be determined \textit{a priori} because a relationship between the
buoyancy flux is unclosed.
One possible closure for this relationship is by using the
laminar-like boundary-layer scaling and the
turbulent bulk scaling as prescribed by the GL theory
(\S\,\ref{subsec:GlobalAveDissip}).
When applied, the buoyancy flux is found to scale as
$\langle u T \rangle_\textit{BL} \sim \Rey^{1/2}$ and
$\langle u^\prime T^\prime \rangle_\textit{bulk} \sim \Rey$
(figure \ref{fig:BuoyIntegralVsRe}), consistent
with the GL prediction.
Hence, to fully extend the GL theory to the present flow,
relationships for the bulk dissipation of the mean,
$\langle \varepsilon_{\overline{u}} \rangle_\textit{bulk}$
and $\langle \varepsilon_{\overline{T}} \rangle_\textit{bulk}$;
mean components influencing the bulk, $U_\textit{bulk}/U$
and $\Delta T_\textit{bulk}/\Delta T$; and mean vertical
buoyancy flux, $\langle \overline{u}\overline{T} \rangle_\textit{bulk}$,
are needed in terms of $\Rey$, $\Ray$, $\Nus$ and $\Pran$.
Current efforts are underway to uncover the aforementioned relationships.
Similar to RB convection, the present results indicate that, for
vertical natural convection, $\Ray$, $\Nus$ and $\Pran$ may be better
related by non-pure power laws that reflect the underlying flow physics.

\section*{Acknowledgements}
The authors gratefully acknowledge the computing time provided by the NCI National
Facility in Canberra, Australia, which is supported by the Australian Commonwealth
Government. Some of the simulations were conducted with the support of iVEC
through the use of advanced computing resources located at iVEC@Murdoch,
Western Australia, Australia.

\bibliographystyle{jfm}

\end{document}